\documentclass[12pt,a4paper]{article}
\usepackage[english]{babel}
\usepackage[latin1]{inputenc}
\usepackage[T1]{fontenc}
\usepackage{pgfplots}
\usepackage{amsmath}
\usepackage{amsfonts}
\usepackage{amssymb}
\usepackage{braket}
\usepackage{graphicx,color}
\usepackage{hyperref}
\usepackage{tikz}
\usepackage{lmodern,bm}
\usepackage{slashed}
\usepackage{dsfont}
\usepackage{float}
\usepackage{pgfplots}
\pgfplotsset{compat=1.18}
\begin{document}
\title{Fermionic dynamical Casimir effect: Magnus
expansion} 
\author{C.~D.~Fosco and G.~Hansen\\
{\normalsize\it Centro At\'omico Bariloche and Instituto Balseiro\/}\\
{\normalsize\it Comisi\'on Nacional de Energ\'{\i}a At\'omica\/}\\
{\normalsize\it R8402AGP S.\ C.\ de Bariloche, Argentina.\/}}
\maketitle
\begin{abstract}
We study pair creation out of the vacuum, for a system consisting of a
massive Dirac field in $1+1$ dimensions, contained between a pair of
perfectly reflecting boundaries, one of them oscillating. After
analyzing some general properties of the vacuum-decay process, we
evaluate the corresponding transition amplitude in a Magnus expansion
of the $S$-matrix. We show how this yields, besides the single-pair
creation amplitude, multipair ones, as well as corrections to the single
pair amplitude.
	
We also apply it to obtain an approximate, yet explicitly unitary
expression for the Bogoliubov transformation between the {\em in\/}
and {\em out\/} Fock spaces.  
\end{abstract}
\section{Introduction}\label{sec:intro}
Startling macroscopic phenomena  arise when quantum fields are subjected
to non-trivial boundary conditions.  Among them, the Casimir
effect~\cite{Casimir:1948}  stands out as a prime example. In its static
version, it manifests itself  in the existence of forces which reflect
the dependence of the vacuum energy  on the boundary conditions.  This
subject of study is the so called {\em static\/} Casimir effect
(SCE)~\cite{Bordag:2001}. On the other hand, the {\em dynamical\/} Casimir 
effect~\cite{Dodonov2020} (DCE),  delves with time-dependent changes in
the boundary conditions. Within the general framework of the DCE, we
consider  here a Dirac field, in a situation where the geometry of the
boundaries  changes, while the (bag model) conditions they impose is
unaltered. 
Under these assumptions, the DCE observables manifest features of the 
space-time geometry swept by the boundaries.
A particularly interesting effect is, under the appropriate
circumstances, the creation of particles out of the vacuum
state. Those particles are real quanta of the field, their nature
being inherited from that of the field upon which the boundary conditions
are imposed. For instance, photons are emitted when the boundaries are
perfect electromagnetic conductors, and fermion pairs when bag
conditions~\cite{Chodos:1974} are imposed upon a Dirac field. The latter
is the kind of system we are  interested in, among other reasons because
it may be relevant in Condensed  Matter Physics, since Dirac fields,
specially in lower dimensions, appear as ingredients of many continuum
models~\cite{Fradkin:2013}.

In~\cite{Mazzitelli:1986qk} the fermionic DCE for a massless 
Dirac field in \mbox{$1+1$} dimensions, satisfying bag conditions on two
moving  boundaries, has been been studied. In~\cite{Fosco:2022} we used a
perturbative  approach, applicable to either massive or massless Dirac
fermions, to study fermion  pair creation  in an oscillating cavity, also
in \mbox{$1+1$} dimensions, to the lowest non-trivial order in the
amplitude of oscillation. It is the aim of this paper to improve upon that
calculation, by performing a different,  yet related expansion: firstly,
we just assume the part of the action taken as perturbation to be 
small (in comparison to the unperturbed action).  This approach keeps 
reparametrization invariance for the curves swept by the boundaries, even 
when working at the lowest non-trivial order in the new expansion. 
By the same token, relativistic corrections are incorporated, since the 
perturbation involves a Lorentz factor.  
Secondly, we use a Magnus expansion to derive transition probabilities,
and interpret its implications from the point of view of the vacuum decay 
amplitudes, and for the mapping between \textit{in} and \textit{out}
Fock spaces.

The paper is structured as follows: in Sect.~\ref{sec:themodel} we define
the model to be considered, briefly reviewing its main aspects. Then, in 
Sect.~\ref{sec:pair}, we discuss some general aspects of the pair creation
phenomenon, and in Sect.~\ref{sec:inter} we deal with the perturbative
evaluation of the decay  amplitude, in the context of the Magnus
expansion. The latter is applied to  obtain an explicitly unitary
approximation for the Bogoliubov transformation, in Sect.~\ref{sec:bogo}.
Finally, in Sect.~\ref{sec:conc} we convey our conclusions.

\section{The model}\label{sec:themodel}
The model we consider here has already been studied, to the lowest
non-trivial order, in~\cite{Fosco:2022} (we refer to that work for the
properties which here are applied without demonstration).  
  The system consists of a Dirac field $(\psi, \bar{\psi})$ in $1+1$
dimensions which is subjected to bag boundary conditions by means of its
coupling to a `potential' $V$, the role of which is to impose the
appropriate boundary conditions of the two borders (see below).  The
real-time action $\mathcal{S}$ for the system is given by
\begin{equation}\label{eq:defs}
	{\mathcal S}(\bar{\psi}, \psi ;V) \;=\; 
	\int d^2x \, \bar{\psi} \, \big( i \not \!\partial  - m - V \big) \,
	\psi \;, 
\end{equation}
where $m$ is the mass of the fermions.  We adopt conventions such that 
Planck's constant $\hbar$ and the speed of light $c$, are both set equal
to $1$. 
Space-time coordinates are denoted by $x^\mu$ (with $\mu = 0, 1$ and
$x^0 = t$), and the Minkowski metric is
$(g_{\mu\nu}) \equiv \text{diag}(1,-1)$.
For Dirac's $\gamma$ matrices, we choose the representation $\gamma^0
\equiv \sigma_1$, $\gamma^1 \equiv i \sigma_3$, where $\sigma_i \, (i =
1,2,3)$ are the standard Pauli's matrices.

Finally, we deal with the form of $V$ which is required in order
to impose the desired boundary conditions, in the approach outlined
in~\cite{Fosco:2008,Ttira:2010}.  The field is assumed to be
in the presence of two walls, labeled by $L$ and $R$, but, 
contrary to what we did in~\cite{Fosco:2022}, we now assume that $L$
moves while $R$ remains static. The first moves along the trajectory
$x^1 = \eta(x^0)$, while $R$ remains at the position $x^1 = a$.
This motion is constrained so that the walls do not overlap, i.e.,
$\eta(x^0)< a$.

Recalling~\cite{Fosco:2022,Fosco:2023}, the potential $V$ for this
scenario assumes the form:
\begin{equation}
	V(x) = g_L \, \gamma^{-1}(u) \, \delta[x^1 - \eta(x^0)] + g_R \,
	\delta(x^1 - a) \, ,
\end{equation}
where $\gamma(u) = 1/\sqrt{1 - u^2}$ denotes the instantaneous Lorentz
factor as a function of $u = u(x^0) \equiv \dot{\eta}(x^0)$. 
The dimensionless constants $g_L$ and $g_R$ determine the nature of 
the resulting boundary conditions: values differing from $2$, result
in imperfect boundary conditions, allowing for the possibility of some
current escaping from, or coming into, the cavity.  Conversely, setting
$g_L = g_R = 2$ establishes~`bag' boundary condition as shown
in~\cite{Fosco:2008}. For the sake of completeness, we summarize those
conditions, for the case at hand. To that end, we introduce the
projectors: $P_L(x^0)=\frac{1 + i \not \hat{n}_L(x^0)}{2}$, for the
moving boundary at $L$, and \mbox{$P_R=\frac{1 + i \not \hat{n}_R}{2}$}, 
for the static one at $R$.
Here, the unit normals are given by:
\begin{equation}
	\hat{n}_L(x^0)\,=\, \gamma(\dot{\eta}(x^0)) \,
	\big( \dot{\eta}(x^0), 1 \big) \;,\;\;\;
	\hat{n}_R \,=\,( 0,  - 1 ) \;, 
\end{equation}
and the boundary conditions become:
\begin{align}
	\label{eq:bag_boundary_conditions}
	P_L(x^0) \, \psi\big(x^0,\eta{(x^0)}^+\big) \,=\, 0 \,=\, P_R \,
	\psi(x^0, a^-)  \;,
\end{align}
where the $+$ and $-$ superscripts indicate `lateral' limits, approaching
the respective point from the right and from the left, respectively.

\section{Pair creation}\label{sec:pair}
We want to compute the probability of vacuum decay, due
to the motion of the $L$ wall. We make the usual assumptions of a finite
temporal extent for the external perturbation of a system, so that both
the initial state (vacuum) and the final state can be characterized in
terms of the states of the Fock space of a free Hamiltonian: the
{\em in\/} and {\em out\/} spaces.
In our case, that Hamiltonian shall be the one corresponding to a
situation where both boundaries are static, and at the same distance
before and after the motion occurs. 

To be more explicit, we split the action as follows:
\begin{align}
	\label{eq:S_action}
	\mathcal{S} = 
	\mathcal{S}_0 + \mathcal{S}_{I} \;,
\end{align}
with
\begin{align}
&\mathcal{S}_{0}(\bar{\psi},\psi) \equiv
	\int d^2x \, \bar{\psi}(x) \, \big( \, i \slashed{\partial} - m
	- V_0(x) \, \big) \, \psi(x) \, ,\\
	&\mathcal{S}_{I}(\bar{\psi},\psi,\eta) \equiv
	-\int d^2x \, \bar{\psi}(x) \, \varphi(x) \, \psi(x) \, ,
\end{align}
and:
\begin{equation}
\begin{aligned}
	V_0(x) &\equiv 2 \, \big[ \delta(x^1) \,+\, \delta(x^1 - a) \big]
	\; , \\
	\varphi(x) &\equiv 2 \, \big[ \gamma^{-1}(\dot{\eta}(x^0)) \, 
	\delta(x^1 - \eta(x^0))\,-\, \delta(x^1) \big] \;.
\end{aligned}
\end{equation}

Note that, contrary to our previous work with similar models, we are not 
expanding the perturbation in powers of the departure $\eta$, keeping
just the lowest (linear) order term. On the contrary, we develop the
expansion keeping the structure for $\varphi$, and in principle also the
$\gamma$ factor which accounts for relativistic corrections.

The free piece, $\mathcal{S}_{0}$, determines the basis of eigenstates of
the Dirac operator with the boundary conditions which result from
(\ref{eq:bag_boundary_conditions}) when $\eta \equiv 0$.  In the case at
hand, this implies finding the solutions to:
\begin{equation}
	(i \slashed{\partial} - m ) \, \psi(x) = 0 \;,
\end{equation}
with the boundary conditions:
\begin{align}
	\frac{1 + i \gamma_1}{2} \, \psi(x^0,0^+) \,=\, 0 \,=\, 
	\frac{1 - i \gamma_1}{2} \, \psi(x^0,a^-) \;.
\end{align}
A convenient basis of solutions to this system is the following:
\begin{equation}
\label{eq:solutions}
\begin{aligned}
	\psi_{p,\pm}(x^1) = \frac{1}{\sqrt{m + a E_p^2}}
	\begin{pmatrix}
		\pm E_p \sin(p x^1) \\ 
		p \cos(p x^1) + m \, \sin(p x^1)
	\end{pmatrix} \, ,
\end{aligned}
\end{equation}
where the $+$ sign corresponds to positive ($E_p$) energy solutions
and $-$ to negative ($-E_p$) energy ones. In both cases
$E_{p} \equiv \sqrt{p^2 + m^2}$ and the values of $p$ are
determined by the transcendental equation:
\begin{equation}
	\label{eq:momenta_eq}
	m a \, \text{sinc}(p a) + \cos(p a) = 0 \, ,
\end{equation}
where the $\text{sinc}$ function is, in our conventions, defined as:
$\text{sinc} \, x = \frac{\sin x}{x}$.

The solutions are orthogonal and normalized as follows:
\begin{equation}
	\int_0^a \, dx^1 \,
	\psi_{p,\sigma}^{\dagger}(x^1) \,
	\psi_{p^\prime,\sigma^\prime}^{\vphantom{\dagger}}(x^1)
	= \delta_{p,p^\prime} \, \delta_{\sigma,\sigma^\prime} \, .
\end{equation}

The spectrum of the system is discrete, and  the interchange
$p \to -p$ does not yield an independent solution.  Thus, it is
sufficient to consider (as we do) $p$ assuming non-negative values only.  

When $m = 0$, the spectrum becomes:
\mbox{$|E_{p_n}| = p_n = (n + \frac{1}{2}) \frac{\pi}{a}$} with
$n = 0, 1, \ldots$ Solutions are labeled by the index $p_n$, which
is related to the particles' spatial momenta when the size of the cavity
tends to infinity, i.e., in the limit $a \to \infty$. For practical
purposes, we will use the term \textit{momenta} to refer specifically to
the index $p_n$.

Both the {\em in\/} and {\em out\/} spaces are isomorphic, and the
associated annihilation
and creation operators correspond to  a  conveniently chosen basis
of solutions to the above equation. For instance, the mode expansion for
the {\em in\/} field operator (we omit the `in' label in both the field
and the operators),
\begin{equation}\label{eq:Psi}
	\psi(x) = \sum_{k} \big[ \, b_k \, e^{-i E_k x^0} \, u_k(x^1) 
	+ \, d_k^{\dagger} \, e^{i E_k x^0} \, v_k(x^1) \, \big] \, ,
\end{equation}
where $u_k(x^1) \equiv \psi_{k,+}(x^1)$ and $v_k(x^1) \equiv
\psi_{k,-}(x^1)$, with $b_{k}^{\dagger}$ and $d_{k}^{\dagger}$ being
creation operators of fermions and anti-fermions, respectively.
 With the normalization we have chosen for the solutions to the Dirac
equation, the only non-vanishing anti-commutators among these operators 
are:
\begin{equation}
\label{eq:anticommutation_relations}
	\{ b_k^{\vphantom{\dagger}} \,,\, b^\dagger_p \} \;=\; \delta_{kp} \;,
	\quad
	\{ d_k^{\vphantom{\dagger}} \,,\, d^\dagger_p \} \;=\; \delta_{kp} \;.
\end{equation}

Also note that a single non-negative discrete label is sufficient to
characterize  either a fermion or anti-fermion state. 

Before actually calculating the decay probability amplitudes, we see that,
since  fermion number is conserved, the vacuum can only decay into one or
more fermion  anti-fermion pairs.
Let us first see how, for the action we are considering, the knowing of
just the  {\em single\/} pair creation amplitude, all the multiple pair
ones can be obtained by Wick's theorem combinatorics and knowedge of the
single pair decay amplitude. 
Indeed, assume that we want to consider a rather general vacuum decay
amplitude, from the vacuum to a certain number, $n$ say, of fermion pairs. 
A rather straightforward calculation shows that the reduction
formulae~\cite{ItzyksonZuber} can be adapted to this system: indeed, the
connected transition amplitudes   are determined by the corresponding
Green's function. In this case ($n$ pairs), it then involves  a number
$n$ of  $\psi$ and $\bar{\psi}$ legs. To obtain the amplitude from such a
Green's function, one acts with the inverse of the free propagator
(in our case: the one in the presence of static walls) on the legs of the
function, and then attaches the  respective free spinor. In other words,
the procedure is rather similar to the one used in order to extract
scattering amplitudes in free space, the difference being in  fact that
the free evolution and wave functions correspond to a static cavity.

Since the full action of the system is
quadratic in the Dirac field, Wick's theorem holds true for that
Green's function: it can be written in terms of combinations invoving
pairings, albeit with a  contraction $S_F$, the exact Dirac
propagator in the presence of a moving boundary: 
\begin{equation}
	S_F(x,y) \;\equiv\; \bra{0} T [\psi(x) \bar{\psi}(y) ] \ket{0} \; ,
\end{equation}
which satisfies:
\begin{equation}
(i \not \! \partial_x - m - V(x)) \, S_F(x,y) = i \, \delta^2(x-y) \, .
\end{equation}

All the contributions to the $n$-pair amplitude can then be written in
terms of that propagator (which, in general, one cannot calculate exactly
in closed form). On the other hand, using the reduction formalism, the
basic amplitude, corresponding to the creation of a fermion anti-fermion
pair, with labels $k$ and $p$, respectively, may be obtained as
follows:
\begin{align}
	& \langle k , p ; {\rm out} | 0 ; {\rm in}\rangle \,=\, 
	\int d^2x \int d^2y \, \bar{u}_k(x^1) \, e^{i E_k x^0} \times \, 
	\nonumber\\
	& (i \not \! \partial_x - m - V_0(x^1)) \, S_F(x,y) \,
	(i \, \overleftarrow{\not \! \partial_y} +  m + V_0(y^1)) \,
	e^{i E_p y^0} \, v_p(y^1) \, .
\end{align}
This allows us to write in a more explicit form the corresponding exact
(yet formal) matrix elements of the transition amplitude $T$, ($S = I
+ i T$, $S$: S-matrix):
\begin{align}
	\label{eq:Tfi}
	& T_{fi} = \int dx^1 \int dy^1 \; \bar{u}_k(x^1) \, M(x^1,y^1) \,
	v_p(y^1) \nonumber\\
	& 
	M(x^1,y^1) = \int dx^0 \int dy^0 \, e^{i (E_k x^0 + E_p y^0)} \, 
	\Big[ \varphi(x) \delta^2(x-y) \,+\, \varphi(x) S_F^{(0)}(x,y)
	\varphi(y) \nonumber\\
	& +\,\int d^2z \varphi(x) S_F^{(0)}(x,z) \varphi(z) S_F^{(0)}(z,y)
	\varphi(y) \,+ \ldots \Big] \, .
\end{align}
Here, we have denoted  the initial and final states by $\ket{i}$ and
$\ket{f}$ respectively. We shall have: $\ket{i} = \ket{0}$ and
$\ket{f} = b_{k\vphantom{p}}^{\dagger\vphantom{p}}
d_{p\vphantom{k}}^{\dagger\vphantom{k}} \ket{0}$, where $b_p^\dagger$
and $d_k^\dagger$ create fermions and
anti-fermions, respectively. $S_F^{(0)}$ is the Dirac propagator for
the fermions in the presence of the static boundaries, which is
determined by the equation:
\begin{equation}
	(i \not \! \partial_x - m - V_0(x)) \, S_F^{(0)}(x,y)
	= i \, \delta^2(x-y) \; .
\end{equation}

Knowing the result for a single pair, it is possible to calculate the
probability amplitude for $n$-pairs. For example, for the creation of
two particle pairs, we consider the states $\ket{i} = \ket{0}$ and
$\ket{f} = b_{k_1\vphantom{p_1}}^{\dagger\vphantom{p}}
d_{p_1\vphantom{k_1}}^{\dagger\vphantom{k}}
b_{k_2\vphantom{p_2}}^{\dagger\vphantom{p}}
d_{p_2\vphantom{k_2}}^{\dagger\vphantom{k}} \ket{0}$. Using the
reduction formula, the probability amplitude for the creation of two 
fermion pairs can be expressed in terms of the single pair amplitude:
\begin{equation}
\begin{aligned}
    \langle k_1 , p_1, k_2, p_2 ; {\rm out} | 0 ; {\rm in}\rangle \,=\,\,
    &\langle k_1 , p_1; {\rm out} | 0 ; {\rm in}\rangle
    \times \langle k_2 , p_2; {\rm out} | 0 ; {\rm in}\rangle \\
    \,+\,\, &\langle k_1 , p_2; {\rm out} | 0 ; {\rm in}\rangle
    \times \langle k_2 , p_1; {\rm out} | 0 ; {\rm in}\rangle \, .
\end{aligned}
\end{equation}

Thus, the transition amplitude for multiple pairs can be systematically
constructed from the single pair amplitude by considering all possible
combinations of pairings. In the next section, we evaluate the
transition probability within the framework of the
interaction representation.

\section{Perturbation theory}\label{sec:inter}
The $S$-matrix may be written as follows:
\begin{equation}
	S = 1 + i \, T = U_I(\infty,-\infty) \, ,
\end{equation}
with $U_I(t_f,t_i)$ denoting the evolution operator, from $t_i$ to $t_f$,
in  the interaction representation, while $T$  is the transition matrix.
The probability of fermion and anti-fermion pair production, will
be denoted as $P_{kp}$, in consonance with the indices characterizing the
respective \textit{momenta} $p$ and $k$.

In the interaction picture, the Hamiltonian for the perturbation becomes:
\begin{equation}\label{eq:HIint}
\begin{aligned}
&\mathcal{H}_I^{\prime}(x^0)
	= 2 \, \gamma^{-1}(x^0) \, \sum_{k,p} \big[
	b_{k\vphantom{p}}^{\dagger}b_{p\vphantom{k}}^{\vphantom{\dagger}}
	\, e^{i (E_k - E_p) x^0} \, g_{kp}(\eta(x^0))
	+ b_{k\vphantom{p}}^{\dagger} d_{p\vphantom{k}}^\dagger
	\, e^{i (E_k + E_p) x^0} \, h_{kp}(\eta(x^0)) \\
	&- d_{k\vphantom{p}} b_{p\vphantom{k}} \, e^{-i (E_k + E_p) x^0}
	\, h_{kp}(\eta(x^0))
	- d_{k\vphantom{p}}^{\vphantom{\dagger}}
	d_{p\vphantom{k}}^\dagger \,
	e^{-i (E_k - E_p) x^0} \, g_{kp}(\eta(x^0)) \,
\big] \; ,
\end{aligned}
\end{equation}
where, for an arbitrary argument $x^1$, the functions appearing above are
given by:
\begin{align}
	g_{kp}(x^1) &\equiv \bar{u}_k(x^1) u_p(x^1)
	= - \bar{v}_k(x^1) v_p(x^1) \; , \\
	h_{kp}(x^1) &\equiv \bar{u}_k(x^1) v_p(x^1)
	= - \bar{v}_k(x^1) u_p(x^1) \, .
\end{align}
One can check that under the transposition of suffixes they satisfy:
\begin{equation}
	\label{eq:gh_functions}
	g_{kp}(x^1) \,=\, g_{pk}(x^1) \; , \;\;\;
	h_{kp}(x^1) \,=\, - h_{pk}(x^1) \, ,
\end{equation}
and that, when $m \equiv 0$, they adopt a rather simple form: 
\begin{align}
	g_{kp}(x^1) &= \frac{1}{a} \, \sin( (k+p) \, x^1) \, , \\
	h_{kp}(x^1) &= \frac{1}{a} \, \sin( (k-p) \, x^1) \,.
\end{align}
For later use, we render also their approximate form for small amplitudes
of motion, namely, for $|\eta(t)| << a$:
\begin{equation}
	g_{kp}(\eta(t)) \, \simeq \frac{1}{a} \, (k+p) \, \eta(t) \; , \;\;
	h_{kp}(\eta(t)) \, \simeq \frac{1}{a} \, (k-p) \, \eta(t) \, .
\end{equation}

To recall the relation between the order in which is calculated the
probability, and the one of the Dyson expansion for the $S$-matrix, given
in terms of the evolution operator, 
\begin{equation}
	S \;=\;	\sum_{n=0}^\infty S^{(n)} \;, \quad
	S^{(n)} \;=\; U_I^{(n)}(+\infty,-\infty) \, ,
\end{equation}
we see that:
\begin{equation}
	P_{kp} \,=\,
	\sum_{n=1}^{\infty} P_{kp}^{(n)} \;,\;\; 
	P_{kp}^{(n)} \,=\,
	\sum_{l=1}^{n-1} {S_{kp}^{(n-l)}}^*\, S_{kp}^{(l)} \;, 
\end{equation}
with $S_{kp}^{(l)}$ denoting the $l^{\rm th}$-order term in the matrix
element of the $S$-matrix.
 
In a previous work~\cite{Fosco:2022}, we dealt with the lowest (second)
order contribution to the probability, which involves just the first
order term in the Dyson expansion. 

Let us now see how the Dyson expansion above is related to the Magnus 
expansion~\cite{Blanes:2010}, which in this case allows for the evolution 
operator to be expressed as an exponential:
\begin{equation}
	U_I(t,t_0) \, = \, \exp \{ - i \,\Omega(t,t_0)\} \, , \quad
	\Omega(t_0,t_0) = 0 \;,
\end{equation}
where the operator $\Omega(t,t_0)$ can be written as a series:
\begin{equation}
\label{eq:Magnus_series}
	\Omega(t,t_0) = \sum_{n = 1}^{\infty} \Omega_n(t,t_0) \;,
\end{equation}
where the first few terms are:
\begin{equation}
\begin{aligned}
\label{eq:Omega_n}
	\Omega_1(t,t_0) &= \int_{t_0}^{t} dt_1 \,
	\mathcal{H}_I^{\prime}(t_1) \, , \\
	\Omega_2(t,t_0) &= \frac{1}{2} \, \int_{t_0}^{t} dt_1 \,
	\int_{t_0}^{t_1} dt_2 \, \left[\mathcal{H}_I^{\prime}(t_1),
	\mathcal{H}_I^{\prime}(t_2)\right] \, , \\
	\Omega_3(t,t_0) &= \frac{1}{6} \, \int_{t_0}^{t} dt_1 \,
	\int_{t_0}^{t_1} dt_2 \, \int_{t_0}^{t_2} dt_3 \,
	\big( \, [\mathcal{H}_I^{\prime}(t_1),
	[\mathcal{H}_I^{\prime}(t_2), \mathcal{H}_I^{\prime}(t_3)] ] \\
	&+ [\mathcal{H}_I^{\prime}(t_3), [\mathcal{H}_I^{\prime}(t_2),
	\mathcal{H}_I^{\prime}(t_1)]] \, \big) \, .
\end{aligned}
\end{equation}
An important aspect of this expansion for $\Omega(t,t_0)$ is its
hermiticity at all orders; namely, unitarity of the evolution operator 
at every order. 
In spite of the previous assertion, of course one can connect the terms
in  both expansions, albeit, as we shall see, there is more content to be
found than making this connection. 
We see that equating the Dyson perturbative series for
$U_I(\infty,-\infty)$ (abbreviated as $U_I$) and the Magnus series, 
\begin{equation}
	U_I
	= 1 + (- i) \, U_I^{(1)} + {(-i)}^2 \, U_I^{(2)} 
	+ \ldots \,
	= \, e^{ -i\,(\Omega_1 + \Omega_2 + \ldots)} \; ,
\end{equation}
where \(\Omega_n \equiv \Omega_n(\infty,-\infty)\), with the first 
terms of the series given by:
\begin{align}
	U_I^{(1)} &= \Omega_1 \, , \\
	\label{eq:UI2} 
	U_I^{(2)} &= \Omega_2 + \frac{1}{2!} \, {(\Omega_1)}^2 \, , \\
	U_I^{(3)} &= \Omega_3 + \frac{1}{2!} \,
	(\Omega_1 \Omega_2 + \Omega_2 \Omega_1)
	+ \frac{1}{3!} \, {(\Omega_1)}^3 \; .
\end{align}

Let us evaluate now $\Omega_1$ and the corresponding contribution to the
transition amplitudes to the first non-trivial order. 
\subsection{First order}\label{sec:first_order}
The first term of the Magnus series (\ref{eq:Magnus_series}) becomes
\begin{align}
	\label{eq:Omega_1}
	\Omega_1
	= \sum_{k,p} \big( \,
	  G_{kp}^{\vphantom{\dagger}} \, b_k^\dagger b_p^{\vphantom{\dagger}}
	- G_{kp}^{*} \, d_k^{\vphantom{\dagger}} d_p^\dagger
	+ H_{kp}^{\vphantom{\dagger}} \,
	  b_{k\vphantom{p}}^{\dagger\vphantom{p}}
	  d_{p\vphantom{k}}^{\dagger\vphantom{k}}
	- H_{kp}^{*} \, d_k^{\vphantom{p\dagger}} b_p^{\vphantom{k\dagger}}
	\, \big) \, ,
\end{align}
where the coefficients $G_{kp}$ and $H_{kp}$ are defined as:
\begin{align}
	\label{eq:GH_coeff}
	G_{kp} &\equiv 2
	\int_{-\infty}^{\infty} dt \, e^{i (E_k-E_p) t} \,
	\gamma^{-1}(\dot{\eta}(t)) \, g_{kp}(\eta(t)) \, , \\
	H_{kp} &\equiv 2 \int_{-\infty}^{\infty} dt \, e^{i (E_k+E_p) t} \,
	\gamma^{-1}(\dot{\eta}(t)) \, h_{kp}(\eta(t)) \, .
\end{align}

Note that $\Omega_1$ inherits the quadratic structure, in creation
and annihilation operators, of the interaction Hamiltonian in
(\ref{eq:HIint}). In other words, the same mononials in those operators
appear. It is rather straightforward to see that that kind of quadratic
form also appears in all the $\Omega_n$: they are also bilinear, and
the same combinations  of operators appear (the only ones which preserve
the fermion number).  Of course the kernels defining the quadratic form
will in general be different for each order.

The operator structure of (\ref{eq:Omega_1}) implies that, at this order,
the only possibility is the creation of a single particle pair, with the
correspoding probability given by:
\begin{equation}
	\label{eq:Pkp(1)}
	P_{kp}^{(1)} = |\braket{f|\Omega_1|i}|^2 = |H_{kp}|^2 \, .
\end{equation}
We note that this result coincides with what we would obtain by
considering the modulus squared of the first-order term in the
expansion of the transition amplitude $T_{fi}$ in (\ref{eq:Tfi}).

Given the symmetry properties (\ref{eq:gh_functions}), the coefficients
$G_{kp}$ are hermitian and $H_{kp}$ are antisymmetric with respect to
index swapping:
\begin{equation}
	G_{kp} = {(G_{pk})}^{*} \, , \quad
	H_{kp} = - H_{pk} \, .
\end{equation}

In scenarios involving small amplitudes and non-relativistic speeds,
i.e., $|\eta(t)| << a$ and $|\dot{\eta}(t)| << 1$, these coefficients
simplify to:
\begin{align}
	G_{pk} &\simeq \frac{2}{a} \, (p+k) \, \tilde{\eta}(p-k) \, ,\\
	H_{pk} &\simeq \frac{2}{a} \, (p-k) \, \tilde{\eta}(p+k) \, ,
\end{align}
where $\tilde{\eta}(\nu)$ denotes the Fourier transform of $\eta(t)$,
with the convention:
$\tilde{\eta}(\nu) \equiv \int dt \, e^{i \nu t} \, \eta(t)$.

As a particular example, for a periodic motion such as:
\begin{equation}
	\label{eq:eta_periodic_motion}
	\eta = \eta_0 \, \sin(\omega t), \quad \eta_0 > 0, \, \omega > 0 \, ,
\end{equation}
under conditions of small amplitude
$\eta_0 << a$ and non-relativistic motion $\omega \eta_0 << 1$, the
coefficients take the form:
\begin{align}
	\label{eq:G_periodic_motion}
	G_{pk} &\simeq \frac{2\pi}{i} \, \frac{\eta_0}{a} \, (p + k) \,
	\big(\delta(k - (p + \omega)) - \delta(k - (p - \omega))\big) \, , \\
	\label{eq:H_periodic_motion}
	H_{pk} &\simeq \frac{2\pi}{i} \, \frac{\eta_0}{a} \, (p - k)
	\, \delta(k - (\omega - p)) \, .
\end{align}
Thus, the first-order probability per unit time for pair creation is:
\begin{equation}
\label{eq:P1_periodic_mov}
\begin{aligned}
	\frac{P_{kp}^{(1)}}{T} \,=\,
	&\frac{2\pi}{a^2} \, \eta_0^2 \, {(k-p)}^2 \,
	\delta(k + p - \omega) \, .
\end{aligned}
\end{equation}
This result indicates that particle production is restricted to
specific frequencies of the boundary motion, and it only occurs when
the oscillation frequency $\omega$ matches the sum of the
\textit{momenta} of the particle pair. Furthermore, due to the discrete
nature of these values, particle production is constrained to frequencies
that satisfy the relation $\omega = \frac{n\pi}{a}$, with
$n = 2, 3, \ldots$.

\begin{figure}[H]\label{fig}
\centering
\includegraphics[clip,width=0.7\columnwidth]{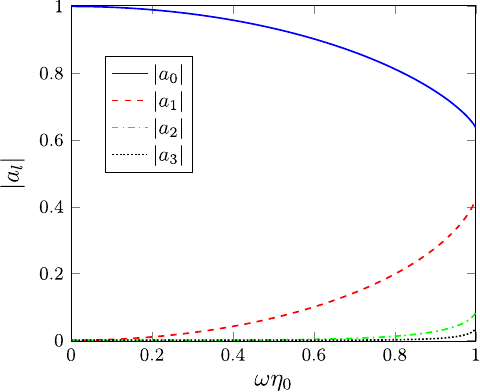}
\caption{Modulus of the Fourier expansion coefficients, $a_l$, as a
function of the mirror's peak velocity, $\omega \eta_0$.}
\end{figure}

Now, let's consider scenarios with small amplitude oscillations where
velocities are not necessarily small. The Lorentz factor, can be
expanded into a Fourier series:
\begin{equation}
	\label{eq:HarmonicMotion}
	\gamma^{-1}(\dot{\eta}(t))
	= \sqrt{1 - \dot{\eta}^2(t)}
	= \sum_{l = 0}^{\infty} a_l \,
	\cos(2 l \, \omega t) \, .
\end{equation}
The coefficients $a_l$ are defined as:
\begin{equation}
a_l = 
\begin{cases} 
    {}_2F_1\left(-\frac{1}{2},\frac{1}{2};1;{(\omega\eta_0)}^2\right) &
	\text{for} \quad l = 0 \, , \\[5pt]
	\frac{2}{l!} \, {(-\frac{1}{2})}_l \, {}_2F_1\left(l-\frac{1}{2},
	l+\frac{1}{2};2l+1;{(\omega\eta_0)}^2\right)
	&\text{for} \quad l \geq 1 \, ,
\end{cases}
\end{equation}
where ${(\cdot)}_{l}$ is the Pochhammer symbol, and ${}_2F_1$ is the
Gaussian hypergeometric function. For this setup, the pair-probability
per unit of time to the first order in the Magnus expansion,  results:
\begin{equation}
\begin{aligned}
	\frac{P_{kp}^{(1)}}{T} \,=\,
	&\frac{2\pi}{a^2} \, \eta_0^2 \, {(k-p)}^2 \, 
	\big[ \, a_0^2 \, \delta(k + p - \omega) \\
	&+ \frac{1}{4} \sum_{l = 1}^{\infty} a_l^2 \,	
	\bigl( \delta(k+p-(2l+1)\,\omega)
	+ \delta(k+p-(2l-1)\,\omega) \bigr)\, \big] \, .
\end{aligned}
\end{equation}
The coefficients $a_l$ represent the weight of the different
contributions to the creation of particles with higher \textit{momenta}
$k$ and $p$. Figure~\ref{fig} displays these coefficients, showing
$|a_l|$ as a function of the peak velocity $\omega\eta_0$. The plot
shows that $a_0 \simeq 1$ and $a_l \simeq 0$ for $l \geq 1$, under small
velocities, aligning with the non-relativistic outcome
(\ref{eq:P1_periodic_mov}).
As velocities increase, higher $a_l$ terms become more significant,
indicating that particles with $k + p \neq \omega$ can be
created.

\subsection{Second order}\label{sec:second_order}
As it is evident in (\ref{eq:UI2}), the second order of the Dyson series,
$U_I^{(2)}$, receives contributions from both the first order term in the
Magnus expansion, $\Omega_1$, and from $\Omega_2$. Let us consider the
first one, which appears squared, and its action on the vacuum:
\begin{equation}
\label{eq:Omega1_squared}
\begin{aligned}
	{(\Omega_1)}^2 \, |0\rangle
	\;=\; 
	& \Big\{\sum_{k,l} \, {[ (G - {\rm tr} \, G) H ]}_{kl} \,
	(b_k^\dagger \, d_l^\dagger + d_k^\dagger \, b_l^\dagger)
	\,+\,
	{\big( \sum_{k,l} b_k^\dagger H_{kl} d_l^\dagger \big)}^2 \\
	&- {\rm tr}(H^{*}H) + {({\rm tr} \, G)}^2 \, \Big\} \, |0\rangle \, ,
\end{aligned}	
\end{equation}
where ${\rm tr}$ denotes the trace. 

Let us see how keeping the exponential form for the $S$ matrix resulting
from the Magnus expansion yields the multipair transition amplitudes, a 
property which, in fact, holds true as a consequence of the quadratic 
structure of the $\Omega_n$ operators-

The key is to identify each contribution to the multipair transition
amplitudes by analyzing the operator structure applied to the vacuum in
(\ref{eq:Omega1_squared}). The first term, with its quadratic structure,
contributes to the creation of one pair, serving as a higher-order
correction to the result in (\ref{eq:Pkp(1)}). 
The second term, characterized by a quartic structure, leads to the
creation of two pairs. Lastly, the term proportional to the identity
accounts for the vacuum-to-vacuum transition amplitude.

On the other and, the second-order term in the Magnus expansion, $\Omega_2$, 
contributes just to the probability amplitude of creating one pair. 
Its contribution is given by:

\begin{align}
	\label{eq:Omega2fi}
	\braket{f|\Omega_2|i}
	= & 2 \sum_{l} \int_{0}^{\infty} dt_1 \,
	\big[ e^{i(E_k+E_l)t_1} \, \left(g_{lp} * h_{kl}\right)(t_1)
	- e^{i(E_k-E_l)t_1} \, \left(h_{lp} * g_{kl}\right)(t_1) \\
	& +  e^{i(E_p+E_l)t_1} \, \left(g_{kl} * h_{lp}\right)(t_1)
	- e^{i(E_p-E_l)t_1} \, \left(h_{kl} * g_{lp}\right)(t_1) 
	\big] \, ,
\end{align}
with the definitions:
\begin{align}
&\begin{aligned}
	\left(g_{lp} * h_{kl}\right)(t_1) \,\equiv\,
	\int_{-\infty}^{\infty} dt_2 \, e^{i(E_k + E_p) t_2} \,
	& \gamma^{-1}(\dot{\eta}(t_2))
	\, \gamma^{-1}(\dot{\eta}(t_1+t_2)) \\
	\, &g_{lp}(t_2) \, h_{kl}(t_1+t_2) \, ,
\end{aligned}
\\
&\begin{aligned}
	\left(g_{kl} * h_{lp}\right)(t_1) \,\equiv\,
	\int_{-\infty}^{\infty} dt_2 \,
	e^{i(E_k + E_p) t_2} \, 
	&\gamma^{-1}(\dot{\eta}(t_2)) \,
	\, \gamma^{-1}(\dot{\eta}(t_1+t_2)) \\
	\, &g_{kl}(t_2) \, h_{lp}(t_1+t_2) \, .
\end{aligned}
\end{align}

As we did with the first order term, we can trace the origin of these
contributions to the one pair creation amplitude back to the terms in the
expansion of the transition amplitude, $T_{fi}$.
In this case, these contributions originate from the second term of the
expansion of $M(x^1,y^1)$ in (\ref{eq:Tfi}).

\section{Bogoliubov transformation}\label{sec:bogo}
In this section, we examine how the dynamic boundary conditions affect
vacuum states and the creation and annihilation operators of the 
fermionic field within the cavity. We specifically focus on how the 
\textit{in} and \textit{out} operators are related, emphasizing the 
transformations induced by changes in boundary conditions.

The \textit{in} vacuum, \( |0\rangle_{\text{in}} \), is the lowest energy 
state defined before the boundaries begin to oscillate. It serves as the 
foundation for the \textit{in} Fock space, which is constructed by the 
successive application of \textit{in} particle and antiparticle creation
operators to  this vacuum state. Similarly, the \textit{out} vacuum,
$|0\rangle_{\text{out}}$, is the lowest energy state after the boundary
motion ceases, and serves as the basis for constructing the \textit{out}
Fock space through similar applications of \textit{out} creation
operators. These two vacuum states are not equivalent, which is
manifested in the possibility of the \textit{in} vacuum evolving into a
state containing particles in the \textit{out} vacuum.

The \textit{in} and \textit{out} operators are connected through
the Bogoliubov transformation $U$:
\begin{equation}
	\begin{pmatrix}
		b \\
		d^\dagger
	\end{pmatrix}_{\text{out}}
	= \, U \, 
	{\begin{pmatrix}
		b \\
		d^\dagger
	\end{pmatrix}}_{\text{in}} ,
\end{equation}
where $U$ is a unitary matrix that preserves the anticommutation
relations between operators in (\ref{eq:anticommutation_relations}),
for both \textit{in} and \textit{out} operators.

To explicitly express the Bogoliubov transformation, we organize the
particle creation and annihilation operators into vectors. The components 
of these vectors are the operators arranged in ascending order of their
\textit{momenta}, as follows:
\begin{equation}
	{(b, d^\dagger)}^T = {(b_{p_0}, b_{p_1}, \ldots, d_{p_0}^\dagger, 
	d_{p_1}^\dagger, \ldots)}^T \, .
\end{equation}
Due to the confined nature of the system, these operators do not
correspond to well-defined momentum states. Instead, they are  linked
to quantized energy states that arise from the perfect reflective~`bag'
boundary conditions specified in equation 
(\ref{eq:bag_boundary_conditions}). Moreover, we emphasize that the
fermions are strictly confined within the cavity  due to these boundary
conditions, and there are no fermions outside of it.

To derive the transformation $U$, we examine a general operator $A$, 
for which we know that the $A_{\text{in}}$ and $A_{\text{out}}$ 
operators are related by a canonical transformation represented by the 
unitary operator $U_I$:
\begin{equation}
	A_{\text{out}} = U_I^{\dagger} \, A_{\text{in}} \, U_I  \, .
	\end{equation}
	At first order in the Magnus expansion, we have:
	\begin{equation}
	\label{eq:A_out}
	A_{\text{out}} \,\simeq\, e^{i \Omega_1} \, A_{\text{in}} \,
	e^{-i \Omega_1} \, .
\end{equation}

Employing Hadamard's formula, we express the right-hand side of
(\ref{eq:A_out}) as an infinite series of nested commutators of
operators $\Omega_1$ and $A_{\text{in}}$:
\begin{equation}
    \label{eq:A_out_series}
    A_{\text{out}}
    \;\simeq\;
	\sum_{n = 0}^{\infty} \frac{i^n}{n!}
    \underbrace{[\Omega_1, \ldots
	[\Omega_1,[\Omega_1}_{\text{n times}},
    A_{\text{in}}]] \ldots ] \, .
\end{equation}

We then introduce a parameter-dependent operator $A(s)$, where
$A(1) = A_{\text{out}}$ and $A(0) = A_{\text{in}}$. Expanding $A(s)$ as
a Taylor series around $s = 0$, yields:
\begin{equation}
	\label{eq:A_series}
	A(s) = \sum_{n=0}^{\infty} \frac{s^n}{n!} \, A^{(n)}(0) \, ,
\end{equation}
where $A^{(n)}$ represents the nth derivative with respect to $s$.
Substituting in (\ref{eq:A_series}) the successive derivatives of the
differential equation:
\begin{equation}
    \label{eq:A_differential_eq}
    i \frac{d}{ds} A(s) = [A(s), \Omega_1] \, ,
\end{equation}
and evaluating at $s = 1$, recovers (\ref{eq:A_out_series}).

Extending this approach to our basis operators,
${(b_k^{\vphantom{\dagger}}(s),d_k^\dagger(s))}^T$, leads to a
differential equation for the operator vector:
\begin{equation}
    i \frac{d}{ds}
    \begin{pmatrix}
        b_k(s) \\
        d_k^\dagger(s)
    \end{pmatrix}
    \;\simeq\;
	\sum_{p}
    M_{kp}
    \begin{pmatrix}
        b_p(s) \\
        d_p^\dagger(s)
    \end{pmatrix},
\end{equation}
where $p$ is summed over the discrete spectrum that results from solving
(\ref{eq:momenta_eq}), and the operator mixing
matrix $M$ is:
\begin{equation}
\label{eq:M_matrix}
M_{kp} \equiv
\begin{pmatrix}
    G_{kp}       & H_{kp}       \\
    -H_{kp}^{*}  & -G_{kp}^{*}
\end{pmatrix}.
\end{equation}

Finally, proceding as we did with $A$ in (\ref{eq:A_series}), we obtain
the Bogoliubov transformation to the first order in the Magnus expansion
for the basis of operators:
\begin{equation}
    \label{eq:Operators_out}
    \begin{pmatrix}
        b \\
        d^\dagger
    \end{pmatrix}_{\text{out}}
    \,\simeq\,
    e^{-i M} \,
    {\begin{pmatrix}
        b \\
        d^\dagger
    \end{pmatrix}}_{\text{in}}.
\end{equation}
We observe that, to this order in the Magnus expansion, particle pair
production requieres the mixing of the $b$ and $d^\dagger$ operators;
therefore, the mixing matrix $M$ must be non-diagonal to enable pair
creation. Consequently, to produce a particle and anti-particle, the
matrix element $H_{kp}$ must be non-zero for some values of $k$ and $p$.

Now, we will calculate the Bogoliubov transformation $U$ for the 
non-relativistic oscillatory movement previously discussed, defined by
equation (\ref{eq:eta_periodic_motion}). For this motion, the values of
$G_{kp}$ and $H_{kp}$ are provided in equations 
(\ref{eq:G_periodic_motion}) and (\ref{eq:H_periodic_motion}). To compute
the transformation, we must evaluate the exponential of the $M$ matrix,
(\ref{eq:M_matrix}). Given that the \textit{momenta} spectrum is
unbounded, the $M$ matrix is infinite dimensional. To derive an explicit
expression for $U$, we introduce a cutoff $p_N$ for the state index. This
results in the vectors ${(b, d^\dagger)}^T$ in (\ref{eq:Operators_out})
becoming finite dimensional:
\begin{equation}
	{(b, d^\dagger)}^T
	=
	{(b_{p_0}^{\vphantom{\dagger}}, b_{p_1}^{\vphantom{\dagger}},
	\ldots, b_{p_N}^{\vphantom{\dagger}},
	d_{p_0}^\dagger, d_{p_1}^\dagger, \ldots, d_{p_N}^\dagger)}^{T} \, ,
\end{equation}
and consequently, leading to a finite dimensional Bogoliubov
transformation matrix.

We note that applying a \textit{momenta} cutoff is equivalent to
considering  a theory with a finite set of fermionic degrees of freedom.
This is reflected in the Fourier representation of the interaction
Hamiltonian, (\ref{eq:HIint}), where the sum over the mixing of
particle and anti-particle creation and destruction operators becomes
finite.

We conclude by presenting results for the Bogoliubov transformation 
$U$ at the threshold frequency $\omega = \frac{2\pi}{a}$, and for 
the two lowest non-trivial cutoffs.

For a cutoff $p_N = p_1 = \frac{\pi}{2a}$, we have:

\begin{equation}
U \,=\, 
\left(
\begin{array}{cccc}
	\cos (\mathcal{N}) & 0 & 0 & \sin (\mathcal{N}) \\
	0 & \cos (\mathcal{N}) & -\sin (\mathcal{N}) & 0 \\
	0 & \sin (\mathcal{N}) & \cos (\mathcal{N}) & 0 \\
	-\sin (\mathcal{N}) & 0 & 0 & \cos (\mathcal{N}) \\
\end{array}
\right) \, ,
\end{equation}
where $\mathcal{N} \equiv \frac{\pi \eta_0 T}{a^2}$, and $T$ is the total
time.

For $p_N = p_2 = \frac{3\pi}{2a}$, we have:
\begin{equation}
U \,=\, 
\left(
\begin{array}{cccccc}
	c & 0 & \frac{3s}{\sqrt{10}}& 0 & \frac{s}{\sqrt{10}} & 0 \\
	0 & \frac{c+9}{10} & 0 & -\frac{s}{\sqrt{10}} & 0 & \frac{3(c-1)}{10} \\
	-\frac{3s}{\sqrt{10}} & 0 & \frac{9c+1}{10} & 0 & \frac{3(c-1)}{10} & 0 \\
	0 & \frac{s}{\sqrt{10}} & 0 & c & 0 & \frac{3s}{\sqrt{10}} \\
	-\frac{s}{\sqrt{10}} & 0 & \frac{3(c-1)}{10} & 0 & \frac{c+9}{10} & 0 \\
	0 & \frac{3(c-1)}{10} & 0 & -\frac{3s}{\sqrt{10}} & 0 & \frac{9c + 1}{10} \\
\end{array}
\right) \, ,
\end{equation}
where $s \equiv \sin\left(\sqrt{10}\,\mathcal{N}\right)$, and
$c \equiv \cos\left(\sqrt{10}\,\mathcal{N}\right)$.

\section{Conclusions}\label{sec:conc}

In this work, we analyzed the dynamical Casimir effect for fermions
confined by moving boundaries in a $1+1$ dimensional setup, with a
particular focus on the role of higher-order perturbative terms and the
Bogoliubov transformation between the creation and annihilation operators.
Our methodology incorporated both the Dyson series and the Magnus
expansion to compute the probability of fermion pair creation due to
boundary oscillations. A notable advantage of the Magnus expansion is its
capability to ensure a unitary evolution of the system at all orders.

We derived formulas for the first and second orders in the Magnus
expansion, specifically calculating the pair creation probabilities at
the first order for both relativistic and non-relativistic periodic
movements. These calculations revealed the dependence of pair creation
probabilities on the speed of the boundary's movements. This analysis
demonstrates how the velocity of boundary movement influences the various
contributions to the probability of pair creation. Furthermore, we
examined how each order contributes not only to the single-pair creation
amplitude but also to multipair creation processes, along with
corrections to the single-pair amplitude.

Additionally, we explicitly computed the Bogoliubov transformation for
the case of non-relativistic periodic movement. This computation further
elucidated the mechanism by which the \textit{in} operators are
transformed into a mixture of \textit{out} operators due to the dynamic
boundary conditions. This shows how quantum states evolve in response to
changes in the geometry and dynamics of the boundary.


\section*{Acknowledgements}
The authors thank ANPCyT, CONICET and UNCuyo for financial support.

\end{document}